\documentclass[preprint,12pt]{elsarticle}

\usepackage{amssymb}
\usepackage{amsmath}
\usepackage{color}
\usepackage{gensymb}
\usepackage[colorlinks=true,citecolor=blue,linkcolor=blue]{hyperref}
\usepackage[dvipsnames]{xcolor}
\usepackage{relsize}

\journal{Nuclear Physics B}

\begin{document}

\begin{frontmatter}



\title{Sterile Neutrino Fits to Short Baseline Data}


\author{G.H. Collin$^a$\footnote{\it email: gabrielc$@$mit.edu},C.A. Arg\"uelles$^{a}$, J.M. Conrad$^a$, M.H. Shaevitz$^b$}

\address{$^{a}$ Massachusetts Institute of Technology, Cambridge, MA 02139, USA}
\address{$^{b}$ Columbia University, New York, NY 10027, USA}

\begin{abstract}
Neutrino oscillation models involving extra mass eigenstates beyond the standard three ($3+N$) are fit to global short baseline experimental data. We find that $3+1$ has a best fit of $\Delta m^2_{41} = 1.75\; \text{eV}^2$ with a $\Delta \chi^2_{null-min}$ (dof) of 52.34 (3). The $3+2$ fit has a $\Delta \chi^2_{null-min}$ (dof) of 56.99 (7). Bayesian credible intervals are shown for the first time for a $3+1$ model. These are found to be in agreement with frequentist intervals.   The results of these new fits favor a higher $\Delta m^2$ value than previous studies, which may have an impact on future sterile neutrino searches such as the Fermilab SBN program.

\end{abstract}


\end{frontmatter}


\section{Introduction}

The well-established discoveries of neutrino mass and 
three-active-flavor mixing can be phenomenologically incorporated into the Standard Model \cite{pdg}, resulting in a model that we can call the ``$\nu$SM''.  This model successfully predicts neutrino oscillations in many experiments.   However,the masses and mixings must be incorporated in an {\it ad hoc} manner. This leads one to ask if there is more ``new physics'' in the neutrino sector that is yet to be discovered that can give us a clearer picture of the underlying theory.

A set of 2$\sigma$ to 4$\sigma$ anomalies have been observed in
short baseline (SBL) oscillation experiments that may indicate new physics.   
SBL experiments have $L/E\sim 1$ m/MeV, where $L$ is the distance from the source to the detector and $E$ is the neutrino energy.
Anomalies are observed from the Liquid Scintillator
Neutrino Detector  (LSND)  experiment\cite{LSND},   the Mini Booster Neutrino
Experiment (MiniBooNE) \cite{miniboonelowe2, nubarminiosc2},  the collection of SBL reactor experiments (often called the ``reactor
Anomaly'') \cite{mention, BUGEY}, and the source calibration data from the gallium-based experiments, SAGE and GALLEX \cite{SAGE3, GALLEX3}.    Any interpretation must also consider similar SBL experiments that have seen no anomalous oscillations (called ``null experiments'') \cite{KARMEN, NuMIMB,  ConradShaevitz, NOMAD1,CCFR84, CDHS, Mahn:2011ea, Cheng:2012yy, MINOSCC1, MINOSCC2}.

Oscillations between active and light sterile neutrinos represent a possible explanation for the combination of anomalous and null SBL data sets.    
Sterile neutrinos are beyond-Standard Model, non-weakly-interacting additions to the neutrino family.
Introducing these new particles extends the number
of mass states and expands the mixing matrix \cite{Conrad:2012qt} in the $\nu$SM.   This allows oscillations with squared mass splittings, $\Delta m^2$, that
are large compared to those in the
$\nu$SM.  Experimental anomalies suggest a mass scale $\sim 1$ eV$^2$. Models  with one
($3+1$), two ($3+2$), and three ($3+3$)  additional sterile neutrino states are generically called ``$3+N$" models.

This paper explores the viable parameter space for oscillation models involving sterile neutrinos.
The most obvious signature of oscillation to sterile neutrinos is disappearance of an active flavor.  Potential $\nu_e \rightarrow \nu_s$ signals have been observed in neutrino and antineutrino mode by the reactor and Gallium-based experiments.   A $\nu_\mu \rightarrow \nu_s$ at a compatible $\Delta m^2$ is yet to be observed, and we will show that this places strong constraints on the phenomenology.  If disappearance occurs, then the model also predicts appearance,  $\nu_\mu \rightarrow \nu_e$ at the same $\Delta m^2$ value(s). This could be consistent with the LSND and MiniBooNE results, which are seen for both neutrinos and antineutrinos.      

This global fit does not make use of the limits from cosmology.  This is because reasonable mechanisms can be put forward within cosmology reduce or remove the constraint, as discussed in Ref. \cite{sterileneutrinowhitepaper}.

 
\section{$3+N$ Fits to Short Baseline Data \label{global}}

The $\nu$SM model has three massive neutrinos leading to two distinct differences between the squared masses, $\Delta m^2_{21}$ and $\Delta m^2_{32}$. The $3 \times 3$ lepton mixing matrix, called the Pontecorvo-Maki-Nakagawa-Sakata (PMNS) matrix, connects the mass eigenstates to the weak interaction eigenstates.

For vacuum oscillations in a $3+N$ model, the probability for finding a neutrino in flavor state $\beta$ after propagating a distance L and being produced as a flavor state $\alpha$ is given\cite{Giunti:2007ry} by
\begin{align}
P_{\alpha \beta} = \delta_{\alpha\beta} &- 4 \sum_{j>i} \text{Re}[ U_{\alpha i}^{*}  U_{\beta i}U_{\alpha j} U_{\beta j}^{*}  ] \sin^{2}\left( \left[\frac{1.27 \;\text{GeV}}{\text{eV}^2\; \text{km}}\right] \frac{\Delta m_{ji}^2 L}{E} \right) \nonumber \\
&+ 2 \sum_{j>i} \text{Im}[ U_{\alpha i}^{*}  U_{\beta i} U_{\alpha j} U_{\beta j}^{*}  ] \sin\left(\left[\frac{2.54 \;\text{GeV}}{\text{eV}^2\; \text{km}}\right] \frac{\Delta m_{ji}^2 L}{E} \right)  \label{oscprob},
\end{align}
where $E$ is the neutrino energy and $\Delta m^2_{ji} = m_j^2 - m_i^2$. Furthermore, the corresponding antineutrino oscillation probability can be obtained by replacing $U\to U^\dagger$. 


\subsection{Incorporating Sterile Neutrinos into the Model}

The 
incorporation of one additional neutrino mass state, in order to extend to a $3+1$ model, introduces a third squared mass splitting.   This also requires an extension of the PMNS matrix to a unitary $4\times4$ matrix:
\begin{equation}
U_{3+1} = \begin{bmatrix}
U_{e1} & U_{e2} & U_{e3} & U_{e4} \\ 
\vdots & & \vdots & U_{\mu4} \\
\vdots & & \vdots & U_{\tau4} \\
U_{s1} & U_{s2} & U_{s3} & U_{s4}
\end{bmatrix}. \label{4mixmx}
\end{equation}
This introduces seven new matrix elements, four of which ($U_{s1},\ldots,U_{s4}$) cannot be directly constrained by experiment due to the non-interacting nature of the fourth `sterile' flavor state. 
The matrix is assumed to be unitary, and the magnitude of the new elements can be constrained by the current measurements of unitarity of the PMNS matrix\cite{Parke:2015goa}.
The new degrees of freedom can be parameterized by introducing three new neutrino mixing angles $\theta_{i4}$ and two new $CP$ violating phases. Eq.~\eqref{oscprob} still holds in describing oscillations, but now the indices $i,j$ run up to 4. 

Although the $3 + 1$ model has three independent squared mass splittings, data indicates that two are small compared to the third. 
The anomalies described in the introduction are all consistent with oscillations corresponding with 
a squared mass splitting on
the order of $1~\text{eV}^2$.
The two splittings associated with the $\nu$SM are  are on the order of $10^{-5}~\text{eV}^2$ and $10^{-3} \text{eV}^2$. The effect of the two small splittings  on an experiment designed to look for $O(1~ \text{eV}^2)$ scale oscillations will be negligible. Therefore, we use the short baseline (SBL) approximation, where we assume that the mass eigenstates that participate in the standard oscillations are degenerate (i.e. $\Delta m^2_{21} = \Delta m^2_{32} = 0$).  

The oscillation probability formula for $\nu_{\alpha} \rightarrow \nu_{\beta}$ in the $3+1$ model then reduces to:
\begin{equation}
P_{\alpha \beta} = \delta_{\alpha\beta} - 4 (\delta_{\alpha\beta} - U_{\alpha 4} U_{\beta 4}^{*}) U_{\alpha 4}^{*} U_{\beta 4} \sin^{2}\left( \left[\frac{1.27 \;\text{GeV}}{\text{eV}^2\; \text{km}}\right] \frac{\Delta m_{41}^2 L}{E} \right). \label{PintermsofU}
\end{equation}
With any particular selection of $\alpha$ and $\beta$ this can be seen to be equivalent to a simple two neutrino model with a mixing amplitude of $\sin^2{2\theta_{\alpha\beta}} = |4(\delta_{\alpha\beta} - U_{\alpha 4} U_{\beta 4}^{*}) U_{\alpha 4}^{*} U_{\beta 4}|$.

More generally, for a $3+N$ model incorporating $N$ sterile neutrinos, the complex phases of $U$ must be taken into account. Let
\begin{equation}
\Phi_{\alpha \beta i j} = \text{arg}(U_{\alpha i} U_{\beta i}^{*} U_{\alpha j}^{*} U_{\beta j} ).
\end{equation}
A transformation of $\nu \rightarrow \bar{\nu}$ causes $\Phi \rightarrow -\Phi$ allowing a difference between neutrino and anti-neutrino oscillations.  These are the $CP$-violating phases. The probability of oscillation for a $3+N$ model can then be written as
\begin{align}
&P(\nu_{\alpha}\rightarrow \nu_{\beta})= \delta_{\alpha\beta} \nonumber \\
 &- 4 \sum_{j>3} \left( \delta_{\alpha\beta} - \sum_{i\geq j} |U_{\alpha i}| |U_{\beta i}| \cos{\Phi_{\alpha\beta i j}} \right) |U_{\alpha j}| |U_{\beta j}| \sin^{2}\left( \left[\frac{1.27 \;\text{GeV}}{\text{eV}^2\; \text{km}}\right]
 \frac{\Delta m_{ij}^2 L}{E} \right) \nonumber  \\  
&+ 2 \sum_{j>3} \left( \delta_{\alpha\beta} -  \sum_{i\geq j} |U_{\alpha i}| |U_{\beta i}|  \sin{\Phi_{\alpha\beta i j}} \right) |U_{\alpha j}| |U_{\beta j}| \sin\left( \left[\frac{2.54 \;\text{GeV}}{\text{eV}^2\; \text{km}}\right]
\frac{\Delta m_{ij}^2 L}{E}\right). 
\end{align}

For $N>1$ sterile neutrinos, the SBL experiments are sensitive to the mass hierarchy through the non-squared sine term. In the global fit, we assume that the degenerate mass states have the lightest mass, {\it i.e.} they follow a normal mass hierarchy.


\subsection{Improved $3+N$ Global Fitting Algorithm \label{algo}}

For this analysis, we have rewritten our previous fitting software \cite{Conrad:2012qt}.   Along with converting from Fortran to C++, 
this package has been designed to make the addition of new data sets easier, as well as to allow the testing of models beyond the $3+N$ presented in this article. 
Also and importantly, we have improved the method of searching the parameter space, which, in our previous fits, did not use a standard  Markov chain Monte Carlo (MCMC) algorithm.   
The new algorithm for searching the parameter space is based on the affine invariant parallel tempering MCMC method used in the Emcee Fitting Package \cite{foreman-mackey_emcee:_2013}. An MCMC efficiently samples the most likely regions of parameter space, whereas a comprehensive scan would be cost-prohibitive.    Technical details of the new approach appear in the appendix to this paper.

The MCMC explores the parameter space by incremental movements governed by the specifics of the algorithm. At each step, a $\chi^2$ value is calculated using the standard definition for normally distributed data:
\begin{equation}
    \chi^2 = \left(\vec{p}(\vec{\theta}) - \vec{d}\right)^T \mathbf{\Sigma} \left(\vec{p}(\vec{\theta}) - \vec{d}\right),
\end{equation}
and a likelihood for Poisson distributed data \cite{baker_clarification_1984}:
\begin{equation}
    \chi^2 = 2 \sum_i^n \left[ p_i(\vec{\theta}) - d_i + d_i \ln\left(\frac{d_i}{p_i(\vec{\theta})}\right) \right],
\end{equation}
where $n$ is the number of bins, $\vec{d}$ the observed data, $\vec{p}(\vec{\theta})$ the model prediction for parameters $\vec{\theta}$, and $\mathbf{\Sigma}$ is the covariance.

These $\chi^2$ values are saved along with their respective $\vec{\theta}$. The algorithm continues until a predetermined number of steps have been executed. From this list, the minimum $\chi^2$ is found. The quantity
\begin{equation}
    \Delta \chi^2(\vec{\theta}) = \chi^2(\vec{\theta}) - \chi^2_{\text{min}},
\end{equation}
is found for each saved $\chi^2$. These $\Delta \chi^2$ values are used to draw the confidence intervals in plots. All points that satisfy
\begin{equation}
    \Delta \chi^2 < \text{CDF}_{\chi^2}^{-1}(k, p),
\end{equation}
are drawn inside the interval with probability $p$.
Where $\text{CDF}_{\chi^2}^{-1}$ is the inverse $\chi^2$ distribution CDF and $k$ is the number of degrees of freedom.
Where there are multiple intervals, they are drawn on the plot in descending order of probability. The plot is effectively a marginalization via minimization. For a 2D plot, the number of degrees of freedom is thus $k=2$.  


\subsection{$3+1$ Frequentist vs. Bayesian Results \label{Bayes}}

In the frequentist treatment (Sec.~\ref{algo}), confidence intervals are drawn from the value of the $\Delta\chi^2$ statistic. For the intervals to be meaningful, the statistic must be correctly $\chi^2$ distributed. This may not necessarily be true, especially in the case of neutrino oscillations where the model predictions use sinusoidal functions. 

Feldman-Cousins\cite{feldman_unified_1998} provides a technique for drawing meaningful confidence intervals in these conditions. However, the method is far too computationally expensive to be used in a global fit. Thus, the frequentist intervals in this paper assume that the $\chi^2$ statistic is correctly distributed.

It would be advantageous to side-step the issue entirely by avoiding the use of a $\chi^2$ statistic. This can be done using Bayesian credible intervals.

For experiments with normally distributed data the log-likelihood is defined using the normal distribution
\begin{equation}
    \ln{\mathcal{L}(\vec{\theta})} = -\frac{1}{2}\left[ (\vec{p}(\vec{\theta}) - \vec{d})^T \mathbf{\Sigma} (\vec{p}(\vec{\theta}) - \vec{d}) + \ln{|\mathbf{\Sigma}|} + n\ln{2\pi} \right],
\end{equation}
and experiments with Poisson distributed data we use,
\begin{equation}
    \ln{\mathcal{L}(\vec{\theta})} = -\sum_i^n \Big[ p_i(\vec{\theta}) - d_i \ln\left(p_i(\vec{\theta}) \right) + \ln{\Gamma(1+d_i)} \Big].
\end{equation}

The density of the explored points in parameter space reflects the underlying posterior distribution $\pi(\vec{\theta})$. An estimate of this posterior is generated from the distribution of walkers with temperature $\beta=1$. Typically a certain number of steps at the beginning of each walker chain contains information about the walkers starting position. As the ensemble begins to equilibrate, this information is lost. The estimate of the posterior should not be polluted by the starting values, so a certain number of steps from the beginning of the chain is typically ignored. These ignored steps are called the ``burn sample."

The $\alpha$ probability credible interval $\mathcal{C}(\alpha)$ must satisfy
\begin{equation}
    \int_{\mathcal{C}(\alpha)} \pi(\vec{\theta}) d\vec{\theta} = \alpha. \label{eq:intervalcondition}
\end{equation}
While there are multiple definitions for $\mathcal{C}$, the most useful when comparing best fits is the highest posterior density interval. Here, the interval is the (possibly disjoint) set of points whose posterior probability meets a threshold $t$:
\begin{equation}
    \mathcal{C} \in \{ \vec{\theta} : \pi(\vec{\theta}) > t \}
\end{equation}
where $t$ is constrained by Eq. \ref{eq:intervalcondition}. Intuitively this can be seen as an interval, which starting at the mode (i.e. the best fit point), grows to include an area whose integrated probability is exactly $\alpha$ and where all points inside the interval have higher probability density than all points outside the interval.

In order to present the Bayesian credible intervals, we plot the highest posterior density interval for a probability $\alpha$,  by drawing the samples whose posterior is greater than a threshold value $t$. The value of $t$ is chosen so that the number of samples meeting this criteria is a fraction $\alpha$ of the total number of samples \cite{chen_monte_1999}.

In both the frequentist and Bayesian cases, the MCMC algorithm was run with a uniform prior on $\log_{10}|U_{ai}|$, $\log_{10}\Delta m_{4i}^2$ and $\log_{10} \Phi$. The positions of the walkers in parameter space was limited as follows:   The matrix elements were required to lie within the space of unitary matrices and be larger than $10^{-6}$.    The phases were required to be less than $2\pi$.  
Large $\Delta m^2$ parameters require much more computing time to evaluate, which slows down the entire ensemble.   Therefore, 
the $\Delta m^2$ parameters are required to be between $10^{-4}\;\text{eV}^2$ and $10^{4}\;\text{eV}^2$ for $3+1$. In the case of additional sterile neutrinos, this was narrowed to $10^{-3}\;\text{eV}^2$ and $10^{3}\;\text{eV}^2$.
Proposed steps outside these listed boundaries are penalized with a log-likelihood of $-\infty$.



\subsection{The Experimental Data Sets \label{expt}}

The full list of experiments included in this study is provided in Tab.~\ref{tab:explist}.
Most data sets used in our past analysis \cite{Conrad:2012qt} have been incorporated into this analysis, however the atmospheric data set and a MiniBooNE disappearance data set that were used previously have been replaced by the MiniBooNE/SciBooNE joint disappearance analyses, which are more restrictive. A second reason to drop the atmospheric constraint was that it assumed no oscillations of electron neutrinos in order to obtain the limit, and this is inconsistent with a global fit.   Also, the description of the LSND experimental result was improved in the code to better represent the published result \cite{LSND}.

The MiniBooNE/SciBooNE data sets in 
neutrino mode\cite{Mahn:2011ea} and anti-neutrino
mode \cite{Cheng:2012yy} were taken from the public release for each analysis.  However, for the neutrino data set,  
an updated covariance matrix was used, along with a cosmic
background data set omitted from the data release \cite{aaaprivate}.  

\begin{table}[!htbp]

\begin{center}

\begin{tabular}{|c|c|c|c|c|}
\hline
Tag & Process  & $\nu$ vs.~$\bar \nu$ & Type & $N_{bins}$ \\ \hline
LSND \cite{LSND}  & $\bar \nu_\mu \rightarrow \bar \nu_e$ & $\bar \nu$ & App & 5\\
KARMEN \cite{KARMEN} & $\bar \nu_\mu \rightarrow \bar \nu_e$ & $\bar \nu$ & App & 9\\
KARMEN/LSND(xsec)  \cite{ConradShaevitz} & $\nu_e \rightarrow \nu_e$ & $\nu$ & Dis & 11\\
BNB-MiniBooNE-$\nu$ \cite{miniboonelowe2, miniboonelowe1} & $\nu_\mu \rightarrow \nu_e$  & $\nu$ &
App & 19\\
BNB-MiniBooNE-$\bar\nu$ \cite{nubarminiosc2, nubarminiosc1}& $\bar \nu_\mu \rightarrow \bar
\nu_e$ & $\bar \nu$ & App & 19\\
NuMI-MB($\nu$app) \cite{NuMIMB} & $\nu_\mu \rightarrow \nu_e$ & $\nu$
& App & 10\\
Bugey \cite{mention, BUGEY}& $\bar \nu_e \rightarrow \bar \nu_e$ & $\bar \nu$ & Dis & 60 \\
Gallium \cite{SAGE3, GALLEX3} & $\nu_e \rightarrow \nu_e$ & $\nu$ & Dis & 4\\
BNB-MiniBooNE/SciBooNE-$\nu$ \cite{Mahn:2011ea}& $\nu_\mu \rightarrow \nu_\mu$ & $\nu$
& Dis & 48\\
BNB-MiniBooNE/SciBooNE-$\bar\nu$ \cite{Cheng:2012yy} & $\bar\nu_\mu \rightarrow \bar\nu_\mu$ & $\bar\nu$
& Dis & 42\\
NOMAD \cite{NOMAD1} & $\nu_\mu \rightarrow \nu_e$  & $\nu$ & App & 30\\
CCFR84 \cite{CCFR84} & $\nu_\mu \rightarrow \nu_\mu$  & $\nu$ & Dis & 18\\
CDHS \cite{CDHS} & $\nu_\mu \rightarrow \nu_\mu$ & $\nu$ & Dis & 15\\
MINOS-CC \cite{MINOSCC1, MINOSCC2}& $\bar \nu_\mu \rightarrow \bar \nu_\mu$ & $\bar \nu$ & Dis & 25\\
\hline
\end{tabular}
\caption{Data sets used in the fits, including the relevant oscillation process, neutrino vs.~antineutrino analyses,  appearance vs.~disappearance analysis, and the number of bins.  See Ref.~\cite{Conrad:2012qt} for more details on each experiment, identified by the ``tag.''}
\label{tab:explist}
\end{center}
\end{table}

\subsection{Updated Fits:  $3+1$\label{sec:results}}

\begin{table}[t]

\begin{center}
\begin{tabular}{|c|cccc|}\hline
 & $N_{\text{bins}}$ & $\chi^2_{min}$ & $\chi^2_{null}$ &  $\Delta \chi^2_{null-min}$ (dof)  \\  \hline\hline
{\bf3+1} &  & & & \ \\  \hline
All & 315 & 306.81 & 359.15 & 52.34 (3)   \ \\  \hline
App & 92 & 88.04 & 150.84 & 62.80 (3)\\  \hline
Dis & 223 & 195.84 & 208.32 & 12.48 (3) \\  \hline
$\nu$ & 155 & 153.18 & 164.57 & 11.39 (3) \\  \hline
$\overline{\nu}$ & 157 & 138.79 & 194.59 & 55.8 (3)  \\  \hline
\hline \hline  

{\bf3+2} & & & & \ \\  \hline
All  & 315 & 302.16 & 359.15 & 56.99 (7) \\  \hline

\end{tabular}
\end{center}

\caption{The $\chi^2$ values, degrees of freedom (dof) and probabilities associated with the best-fit and null hypothesis in each scenario.  $P_{best}$ is the $\chi^2$-probability at the best fit point and $P_{null}$ is the  $\chi^2$-probability at null (no oscillation). }
\label{tab:fitstats}
\end{table}

\begin{table}

\begin{center}
\begin{tabular}{|c|ccc|}\hline
{\bf 3+1} & $\Delta m^2_{41}$ & $|U_{e 4}|$ & $|U_{\mu 4}|$   \ \\  \hline\hline 
All & 1.75 & 0.163 & 0.117 \\  \hline
App & 4.75$\times 10^{-2}$ & 0.743 & 0.638 \\  \hline
Dis & 7.79 & 0.217 & 2.94$\times 10^{-2}$ \\  \hline
$\nu$ & 7.71 & 0.248 & 5.67$\times 10^{-2}$  \\  \hline
$\overline{\nu}$& 5.73 & 0.199 & 0.140    \ \\  \hline 
\end{tabular}
\vspace*{5mm}

\begin{tabular}{|c|ccccccc|}\hline

{\bf 3+2} & $\Delta m^2_{41}$ & $\Delta m^2_{51}$ & $|U_{e 4}|$  & $|U_{\mu 4}|$ & $|U_{e 5}|$ & $|U_{\mu 5}|$ &
$\phi_{54}$  \ \\  \hline\hline
All & 0.475 & 0.861 & 0.120 & 0.177 & 0.141 & 0.111 & $0.0662\pi$ \\  \hline
\end{tabular}
\end{center}

\caption{The oscillation parameter best-fit points in each scenario considered.  The values of  $\Delta m^2$ shown are in units of eV$^2$}
\label{tab:bfpoints}
\end{table}

\begin{figure}[tbp!]
\includegraphics[width=0.48\columnwidth]{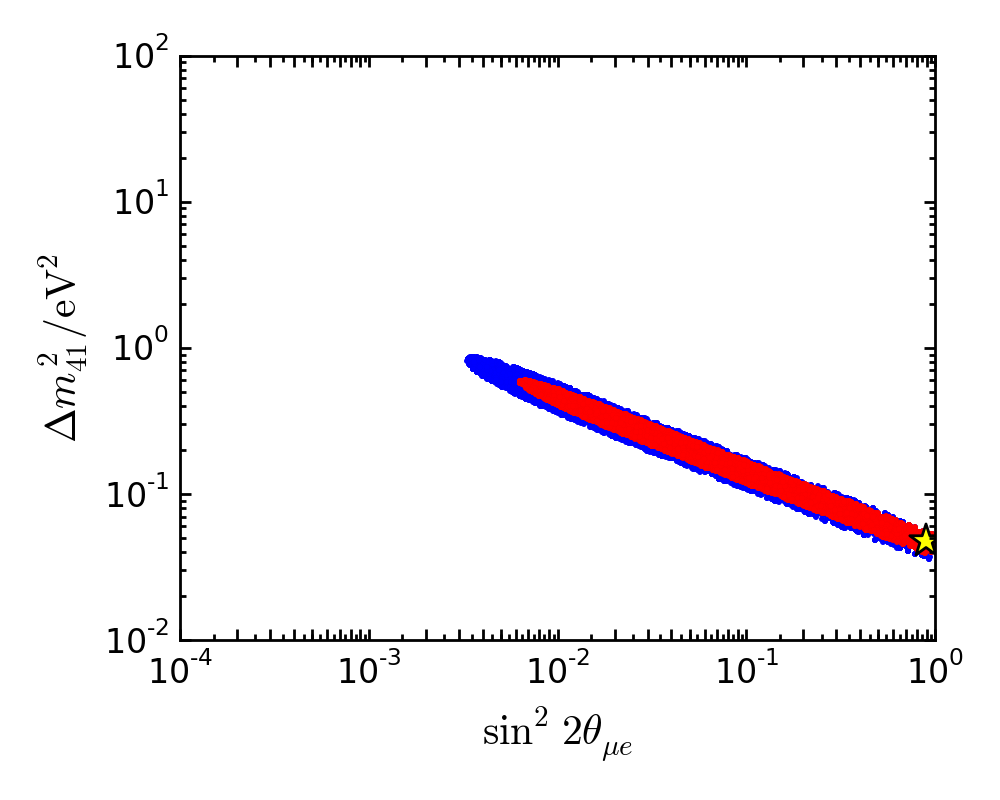}~
\includegraphics[width=0.48\columnwidth]{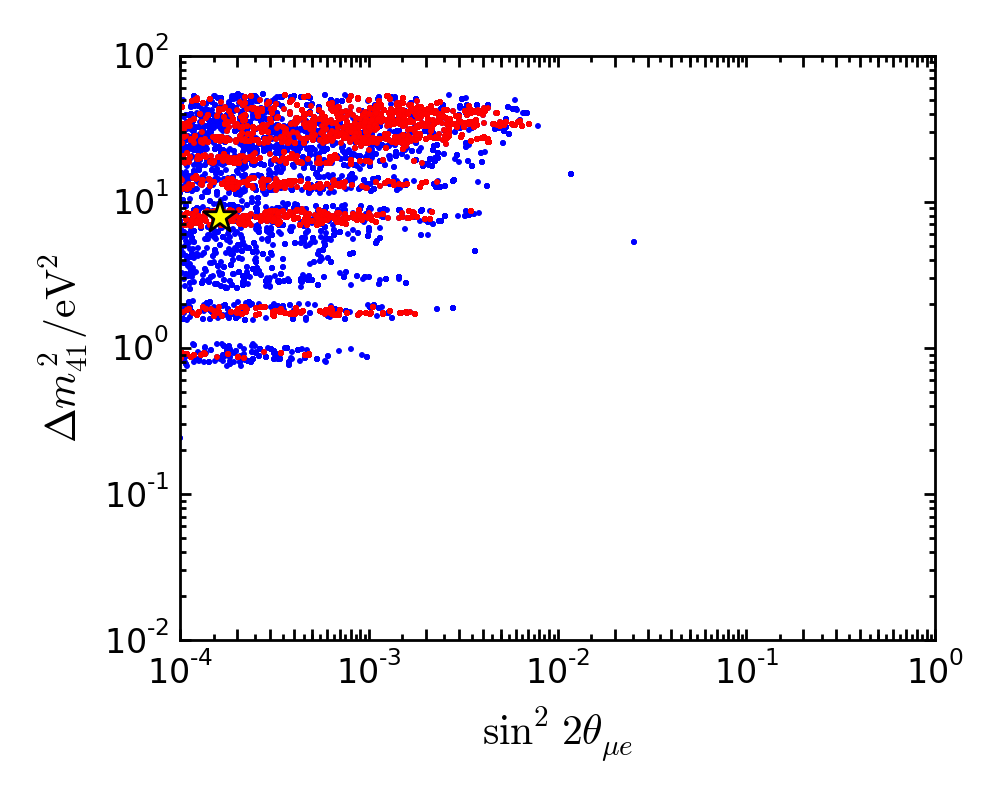} \\
\includegraphics[width=0.48\columnwidth]{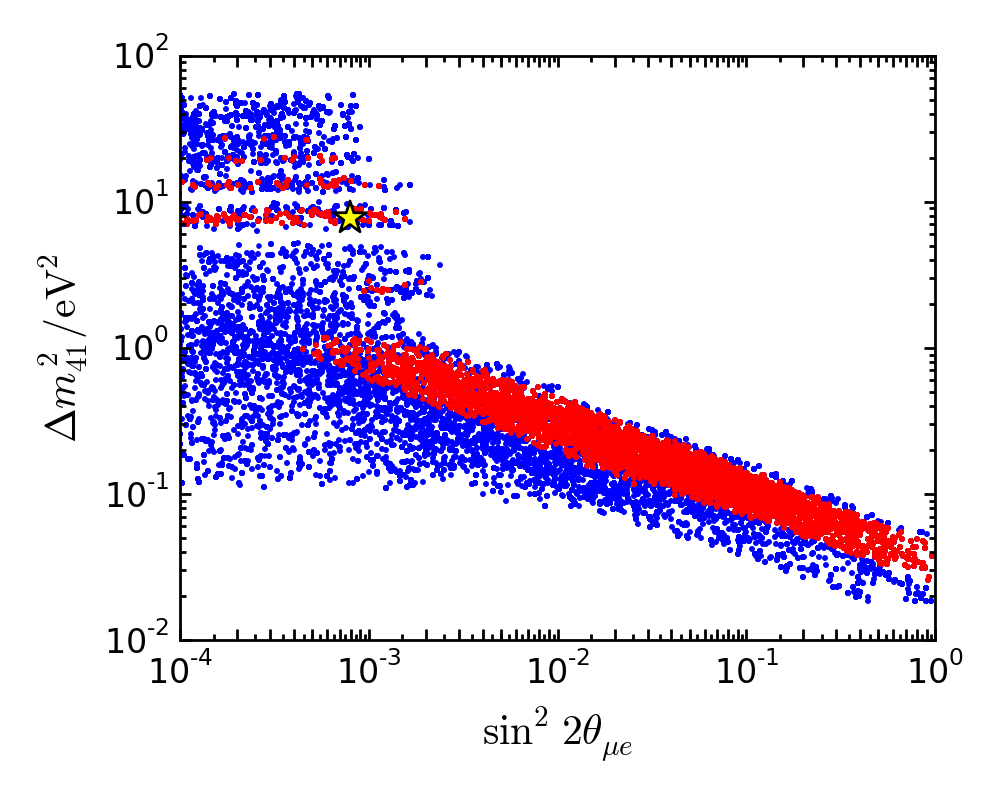}~
\includegraphics[width=0.48\columnwidth]{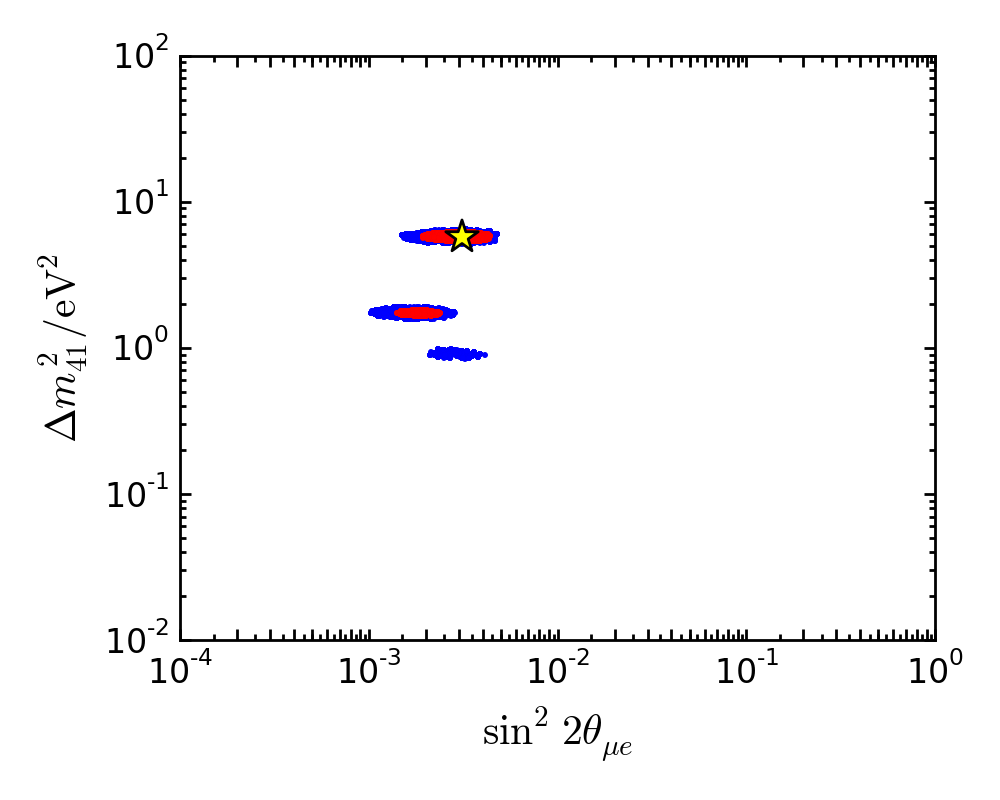} \\
\includegraphics[width=0.48\columnwidth]{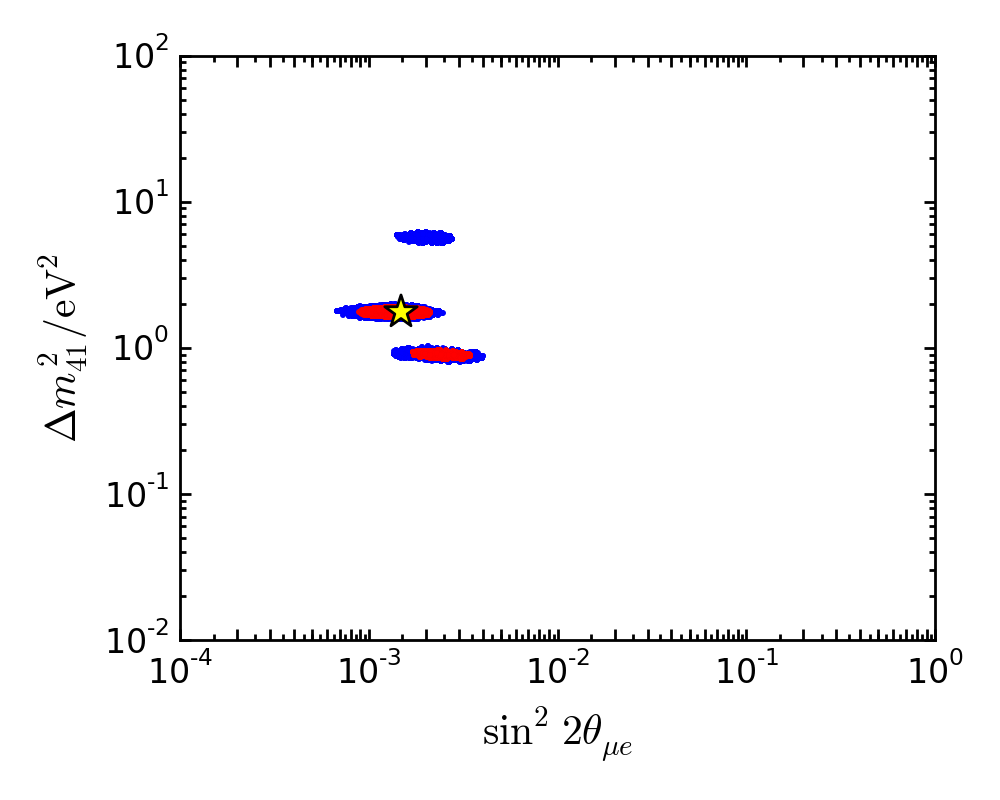}~
\includegraphics[width=0.48\columnwidth]{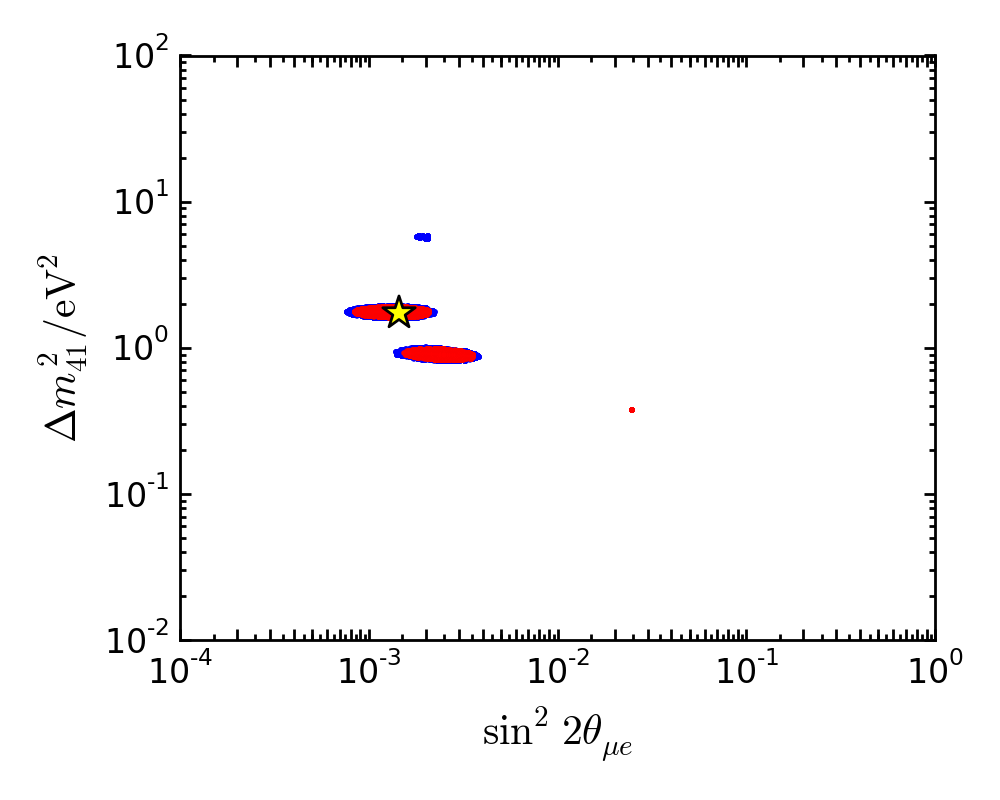}\\
\caption{Frequentist confidence intervals for a $3+1$ model using appearance only data (top left), disappearance data (top right), neutrino data (mid left), anti-neutrino data (mid right), and global data (bottom left).  The Bayesian credible intervals for $3+1$ global data are shown bottom right.
In these plots, $\sin^2 2\theta_{\mu e} = 4 |U_{e4}|^2 |U_{\mu 4}|^2$.   Red indicates 90\% CL and blue indicates 99\% CL. \label{fig:fitresults}}

\end{figure}

Confidence intervals for the frequentist fits to a $3+1$ model are shown in Fig.~\ref{fig:fitresults}, top, middle and bottom left.
The top row shows fits for appearance ($\nu_\mu \rightarrow \nu_e$) and disappearance (muon and electron flavor) disappearance data sets separately, presented on the $\sin^2 2\theta_{\mu e}$ {\it vs.} $\Delta m^2$ plane.   Note that there is no overlap between the 90\% (red) or 99\% (blue) confidence level (CL) regions when the data sets are divided in this manner.   Thus, there is clearly tension between appearance and disappearance experiments.   
The middle row shows the neutrino (left) and antineutrino (right) data sets fit separately within a $3+1$ model.   Dividing the data in this manner, there is overlap between the two data sets, however the antineutrino data sets are highly restrictive.   The global fit for all data sets is shown on the bottom left.   
The quality of the fits is described in Tab.~\ref{tab:fitstats} and 
parameters of the best fit points are provided in Tab.~\ref{tab:bfpoints}.

The $3+1$ global fit in Fig.~\ref{fig:fitresults} have two 90\% allowed regions. This is in contrast to the single 90\% allowed region shown in Ref.~\cite{Conrad:2012qt}. Both share a region at $\sim 1$ eV$^2$, while the new result has a region at $\sim 1.7$ eV$^2$. This new region is a consequence of the improved description of LSND. 

The best fit has moved to the $\sim 1.7$ eV$^2$ region in the new result. This was caused by the addition of the SciBooNE/MiniBooNE disappearance analyses. The changes made to the datasets is discussed in Sec~\ref{expt}.

The credible intervals of the Bayesian fit are shown in the bottom right of Fig.~\ref{fig:fitresults}. The 90\% Bayesian credible intervals are compatible with the 90\% frequentist confidence intervals shown in the bottom left plot.
We note slightly worse agreement in the 99\% credible and confidence intervals of these plots, where the $\Delta m^2\approx 5$ interval is substantially smaller in the Bayesian result. 

\subsection{Updated Fits:  $3+2$\label{sec:results2}}

\begin{figure}[tbp!]
\includegraphics[width=0.48\columnwidth]{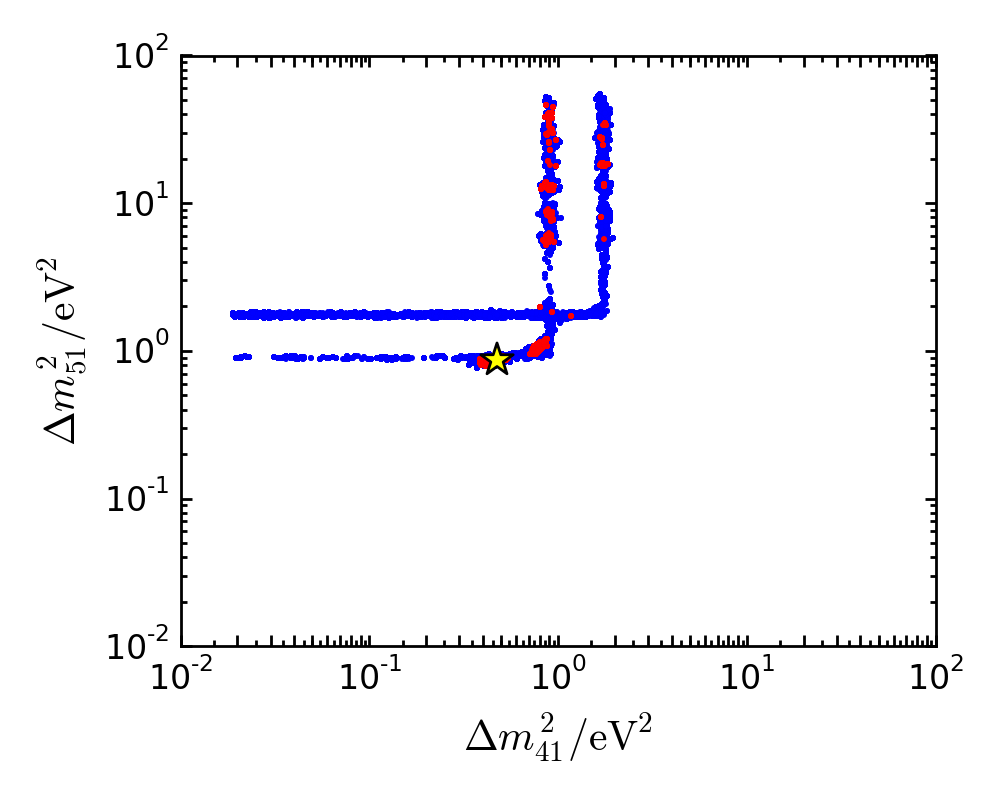}~
\includegraphics[width=0.48\columnwidth]{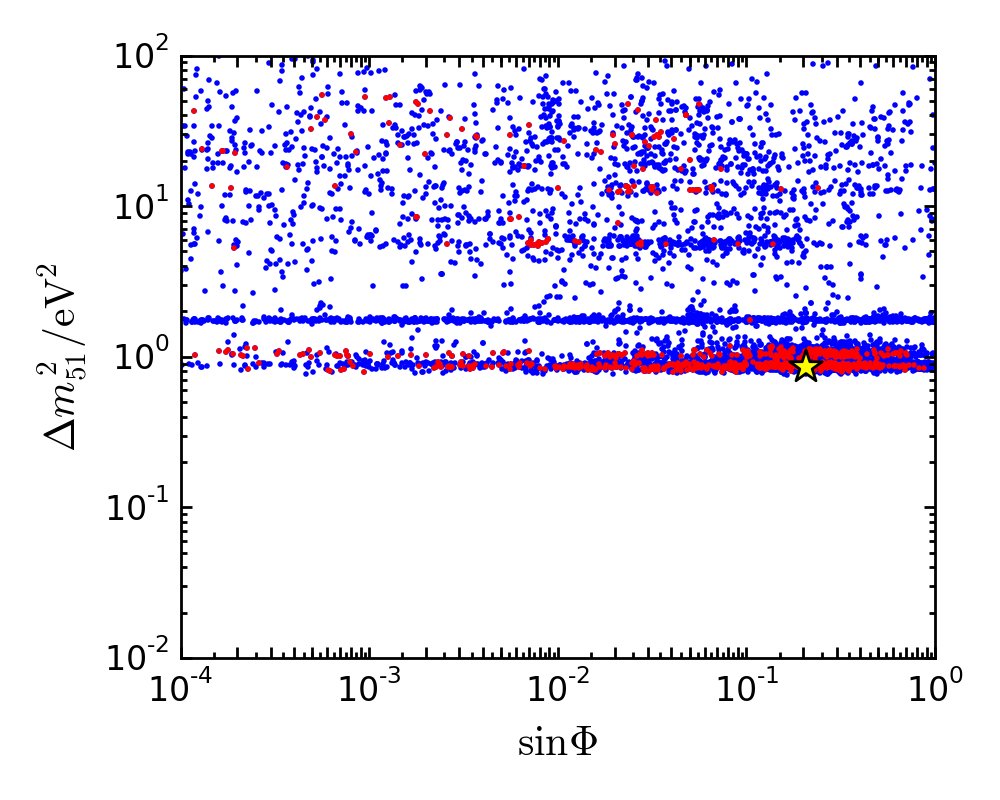}
\caption{Frequentist confidence intervals for a $3+2$ model using global data.  Left:  The parameter space projected into the plane of the two mass splittings.   Right: $\Delta m^2_{51}$ {\it vs.} the $CP$ violation parameter, $\Phi$.  Red indicates 90\% CL and blue indicates 99\% CL.  \label{fig:fitresults2}}

\end{figure}

To relieve the tension in the $3+1$ model, one can move to a $3+2$ model.   The frequentist global fit for this result is shown in Fig.~\ref{fig:fitresults2}.   This model has 7 parameters, and so we select some examples to illustrate the allowed parameter space. 
Fig.~\ref{fig:fitresults2}, left, shows the space of the two mass splittings.  The best fit is for the solution where both splittings are less than 1 eV$^2$ (see Tab. \ref{tab:bfpoints}).  
However, one can see that in the region of $\Delta m^2_{41} \sim 1\;\text{eV}^2$, there are multiple high $\Delta m^2$ solutions that have roughly the same $\chi^2$ value.   
Thus, while our new fit appears at first glance to be a dramatic change from Ref.~\cite{Conrad:2012qt}, which found best fit values of $3+2$ of $\Delta m_{41}^2=0.92\; \text{eV}^2$ and $\Delta m_{51}^2= 17\;\text{eV}^2$, in fact this is actually a small shift of $\chi^2$. 
The previous best fit from Ref. \cite{Conrad:2012qt} remains within the allowed region.  

Fig.~\ref{fig:fitresults2}, right, shows the value of $\Delta m_{51}^2$ as a function of the $CP$ violating parameter.  This shows that the $CP$ violation parameter can shift over a wide range to accommodate many ($\Delta m_{41}^2$, $\Delta m_{51}^2$) pairs of solutions.   Introducing the $CP$ parameter does not greatly improve the fit, however.   As can be seen from Tab.~\ref{tab:fitstats}, the difference in $\Delta \chi^2_{null-min}$ for $3+1$ versus $3+2$ models is about four, while four degrees of freedom were added.

\section{Summary and discussion}

Using the improved software package, we have presented two new results.  First, in a global analysis of the SBL data, we find that a $3+1$ model has a best fit of $\Delta m^2_{41} = 1.75\; \text{eV}^2$ with a $\Delta \chi^2$ (dof) of 52.34 (3) with respect to the null hypothesis.   Second, for the first time we have demonstrated that our fit results are stable if one uses a frequentist or a Bayesian approach.   

The fact that our new fits favor a $\sim$ 2 eV$^2$ solution has interesting implications for the immediate future of sterile neutrino studies.
MicroBooNE \cite{SBNproposal}, which has just begun to take data,  
is located on the Booster Neutrino Beamline (BNB) with a peak $\nu_\mu$ energy of 700 MeV.  The 170 t detector is located at 470 m from the BNB target.  MicroBooNE is directly upstream of the 800 t MiniBooNE experiment, which is at 540 m from the BNB target. If the 2 eV$^2$ solution of a $3+1$ model is correct, then 
MicroBooNE sits closer to oscillation maximum than MiniBooNE, thus predicting a higher signal in MicroBooNE than simple scaling for solid angle and tonnage assumes. On the other hand, 
the ICARUS T600 detector, planned for 600 m from the BNB target \cite{SBNproposal},  may be poorly located to address this 2 eV$^2$ solution.   However, the combination of the three SBN detectors \cite{SBNproposal} including SBND, MicroBooNE, and ICARUS should be able to cover the full range of interest for a 3+1 sterile neutrino signal, given sufficient statistics.

\section*{Appendix: Implementation of MCMC}

The fitting algorithm used in this study is based on the affine invariant parallel tempering Markov chain Monte Carlo (MCMC) method used in the Emcee fitting package \cite{foreman-mackey_emcee:_2013}. An MCMC moves randomly in the parameter space. Each movement is called a step. Before a new step is entered into the history of the Markov chain, it must first pass a probabilistic test. The acceptance probability is based on a Boltzmann distribution:
\begin{equation}
    e^{-E(\vec{\theta})},
\end{equation}
where $E$ is the energy of a position $\vec{\theta}$ in parameter space. This energy is a function of the log-likelihood of the posterior $\pi(\vec{\theta})$
\begin{equation}
    E(\vec{\theta}) = f(\ln{\mathcal{L}(\vec{\theta})}).
\end{equation}
With suitable definitions, the log-likelihood can be related to the $\chi^2$ by
\begin{equation}
    \ln{\mathcal{L}(\vec{\theta})} = - \frac{1}{2} \chi^2(\vec{\theta}).
\end{equation}

A set of $N$ seed points are selected randomly in logarithmic parameter space according to a uniform distribution. Each seed is the beginning of an independent Markov chain called a `walker'. Collections of these walkers are arranged in groups called ensembles.

The walkers are evolved in a step-wise fashion. At each step, the affine invariant movement algorithm is performed on each walker, followed by the parallel tempering swap.
In traditional Metropolis-Hastings movement the new walker location is chosen based on a multi-variate normal distribution. The parameters of this distribution need to be chosen in advance. If the shape of the distribution does not resemble the underlying posterior then inefficient sampling will result.
In comparison, the affine invariant method\cite{goodman_ensemble_2010} only requires the affine scale $a$ to be chosen in advance. The movement of the walkers is based on the current ensemble. Hence, any affine transformation of a normal distribution will be efficiently sampled.

For a given walker ($i$), the affine invariant movement randomly selects another walker ($j$) in the ensemble and attempts to move toward it. The proposed new set of parameters at step $n+1$ is
\begin{equation}
    \vec{\theta}_i(\text{proposed}) = \vec{\theta}_j(n) + z \;[\vec{\theta}_j(n) - \vec{\theta}_i(n)]
\end{equation}
Where $\vec{\theta}_i(n)$ is the parameters of walker $i$ at step $n$ and $z$ is a step distance which is randomly selected according to the distribution
\begin{equation}
    \text{PDF}(z) = \begin{cases} \frac{1}{\sqrt{z}} & \frac{1}{a} < z < a \\ 0 & \text{otherwise} \end{cases}
\end{equation}
Here $a$ is called the affine scale and is set to $2$.
The new set of parameters are then accepted according to the probability
\begin{equation}
    \text{min}\left[1, \: z^{k-1} \frac{e^{-E_i(\vec{\theta}_i(\text{proposed}))} }{ e^{-E_i(\vec{\theta}_i(n))} } \right],
\end{equation}
where $k$ the number of parameters in the model.

The affine invariant method has problems sampling multi-modal distributions. Parallel tempering is a well known MCMC method for sampling multi-modal posterior distributions\cite{earl_parallel_2005}. 
Multiple ensembles of walkers are evolved in parallel. Each of these ensembles has its own temperate parameter $T = 1/\beta$. The energy function for the walker is then defined as
\begin{equation}
    E_i(\vec{\theta}) = \beta_i \ln{\mathcal{L}(\vec{\theta})}.
\end{equation}
This ``flattens'' the 
posterior distribution for ensembles with a large temperate parameter. Walkers in these ensembles have an easier time moving out of a local maximum and exploring the space for other potential maxima.

The information from these high temperature ensembles needs to be communicated back to the low temperature ensembles so that they can be sampled. This is achieved by occasionally swapping the positions of walkers between ensembles.
On each step, a swap is performed with probability $\theta = 0.1$. Random pairs $(i,j)$ of walkers are selected, with walkers in different ensembles. The walkers then swap position with probability
\begin{equation}
    \text{min}\left[ 1, \: \frac{ e^{-E_i(\vec{\theta}_j(n))} }{ e^{-E_i(\vec{\theta}_i(n))} } \frac{ e^{-E_j(\vec{\theta}_i(n))} }{ e^{-E_j(\vec{\theta}_j(n))} } \right].
\end{equation}


\section*{Acknowledgements}
GC, CA and JC are supported by NSF grants 1505858 and 1505855, and MS is  supported  by  NSF  grant  1404209. We thank Christina Ignarra, Benjamin Jones, William Louis, and Jordi Salvado for useful discussion.  We thank Maxim Goncharov for computing support.

\newpage

\bibliographystyle{apsrev}
\bibliography{nus_fit}

\begin{thebibliography}{32}
\expandafter\ifx\csname natexlab\endcsname\relax\def\natexlab#1{#1}\fi
\expandafter\ifx\csname bibnamefont\endcsname\relax
  \def\bibnamefont#1{#1}\fi
\expandafter\ifx\csname bibfnamefont\endcsname\relax
  \def\bibfnamefont#1{#1}\fi
\expandafter\ifx\csname citenamefont\endcsname\relax
  \def\citenamefont#1{#1}\fi
\expandafter\ifx\csname url\endcsname\relax
  \def\url#1{\texttt{#1}}\fi
\expandafter\ifx\csname urlprefix\endcsname\relax\def\urlprefix{URL }\fi
\providecommand{\bibinfo}[2]{#2}
\providecommand{\eprint}[2][]{\url{#2}}

\bibitem[{\citenamefont{Olive et~al.}(2014)}]{pdg}
\bibinfo{author}{\bibfnamefont{K.~A.} \bibnamefont{Olive}} \bibnamefont{et~al.}
  (\bibinfo{collaboration}{Particle Data Group}), \bibinfo{journal}{Chin.
  Phys.} \textbf{\bibinfo{volume}{C38}}, \bibinfo{pages}{090001}
  (\bibinfo{year}{2014}).

\bibitem[{\citenamefont{Aguilar-Arevalo et~al.}(2001)}]{LSND}
\bibinfo{author}{\bibfnamefont{A.}~\bibnamefont{Aguilar-Arevalo}}
  \bibnamefont{et~al.} (\bibinfo{collaboration}{LSND Collaboration}),
  \bibinfo{journal}{Phys. Rev. D} \textbf{\bibinfo{volume}{64}},
  \bibinfo{pages}{112007} (\bibinfo{year}{2001}).

\bibitem[{\citenamefont{Aguilar-Arevalo et~al.}(2009)}]{miniboonelowe2}
\bibinfo{author}{\bibfnamefont{A.~A.} \bibnamefont{Aguilar-Arevalo}}
  \bibnamefont{et~al.} (\bibinfo{collaboration}{MiniBooNE}),
  \bibinfo{journal}{Phys. Rev. Lett.} \textbf{\bibinfo{volume}{102}},
  \bibinfo{pages}{101802} (\bibinfo{year}{2009}), \eprint{0812.2243}.

\bibitem[{\citenamefont{Aguilar-Arevalo et~al.}(2013)}]{nubarminiosc2}
\bibinfo{author}{\bibfnamefont{A.~A.} \bibnamefont{Aguilar-Arevalo}}
  \bibnamefont{et~al.} (\bibinfo{collaboration}{MiniBooNE}),
  \bibinfo{journal}{Phys. Rev. Lett.} \textbf{\bibinfo{volume}{110}},
  \bibinfo{pages}{161801} (\bibinfo{year}{2013}), \eprint{1207.4809}.

\bibitem[{\citenamefont{Mention et~al.}(2011)\citenamefont{Mention, Fechner,
  Lasserre, Mueller, Lhuillier et~al.}}]{mention}
\bibinfo{author}{\bibfnamefont{G.}~\bibnamefont{Mention}},
  \bibinfo{author}{\bibfnamefont{M.}~\bibnamefont{Fechner}},
  \bibinfo{author}{\bibfnamefont{T.}~\bibnamefont{Lasserre}},
  \bibinfo{author}{\bibfnamefont{T.}~\bibnamefont{Mueller}},
  \bibinfo{author}{\bibfnamefont{D.}~\bibnamefont{Lhuillier}},
  \bibnamefont{et~al.}, \bibinfo{journal}{Phys. Rev. D}
  \textbf{\bibinfo{volume}{83}}, \bibinfo{pages}{073006}
  (\bibinfo{year}{2011}).

\bibitem[{\citenamefont{Declais et~al.}(1995)\citenamefont{Declais, Favier,
  Metref, Pessard, Achkar et~al.}}]{BUGEY}
\bibinfo{author}{\bibfnamefont{Y.}~\bibnamefont{Declais}},
  \bibinfo{author}{\bibfnamefont{J.}~\bibnamefont{Favier}},
  \bibinfo{author}{\bibfnamefont{A.}~\bibnamefont{Metref}},
  \bibinfo{author}{\bibfnamefont{H.}~\bibnamefont{Pessard}},
  \bibinfo{author}{\bibfnamefont{B.}~\bibnamefont{Achkar}},
  \bibnamefont{et~al.}, \bibinfo{journal}{Nucl. Phys. B}
  \textbf{\bibinfo{volume}{434}}, \bibinfo{pages}{503} (\bibinfo{year}{1995}).

\bibitem[{\citenamefont{Abdurashitov et~al.}(2009)}]{SAGE3}
\bibinfo{author}{\bibfnamefont{J.}~\bibnamefont{Abdurashitov}}
  \bibnamefont{et~al.} (\bibinfo{collaboration}{SAGE Collaboration}),
  \bibinfo{journal}{Phys. Rev. C} \textbf{\bibinfo{volume}{80}},
  \bibinfo{pages}{015807} (\bibinfo{year}{2009}).

\bibitem[{\citenamefont{Kaether et~al.}(2010)\citenamefont{Kaether, Hampel,
  Heusser, Kiko, and Kirsten}}]{GALLEX3}
\bibinfo{author}{\bibfnamefont{F.}~\bibnamefont{Kaether}},
  \bibinfo{author}{\bibfnamefont{W.}~\bibnamefont{Hampel}},
  \bibinfo{author}{\bibfnamefont{G.}~\bibnamefont{Heusser}},
  \bibinfo{author}{\bibfnamefont{J.}~\bibnamefont{Kiko}}, \bibnamefont{and}
  \bibinfo{author}{\bibfnamefont{T.}~\bibnamefont{Kirsten}},
  \bibinfo{journal}{Phys. Lett. B} \textbf{\bibinfo{volume}{685}},
  \bibinfo{pages}{47} (\bibinfo{year}{2010}).

\bibitem[{\citenamefont{Armbruster et~al.}(2002)}]{KARMEN}
\bibinfo{author}{\bibfnamefont{B.}~\bibnamefont{Armbruster}}
  \bibnamefont{et~al.} (\bibinfo{collaboration}{KARMEN Collaboration}),
  \bibinfo{journal}{Phys. Rev. D} \textbf{\bibinfo{volume}{65}},
  \bibinfo{pages}{112001} (\bibinfo{year}{2002}).

\bibitem[{\citenamefont{Adamson et~al.}(2009)}]{NuMIMB}
\bibinfo{author}{\bibfnamefont{P.}~\bibnamefont{Adamson}} \bibnamefont{et~al.}
  (\bibinfo{collaboration}{MiniBooNE, MINOS}), \bibinfo{journal}{Phys. Rev.
  Lett.} \textbf{\bibinfo{volume}{102}}, \bibinfo{pages}{211801}
  (\bibinfo{year}{2009}), \eprint{0809.2447}.

\bibitem[{\citenamefont{Conrad and Shaevitz}(2012)}]{ConradShaevitz}
\bibinfo{author}{\bibfnamefont{J.}~\bibnamefont{Conrad}} \bibnamefont{and}
  \bibinfo{author}{\bibfnamefont{M.}~\bibnamefont{Shaevitz}},
  \bibinfo{journal}{Phys. Rev. D} \textbf{\bibinfo{volume}{85}},
  \bibinfo{pages}{013017} (\bibinfo{year}{2012}).

\bibitem[{\citenamefont{Astier et~al.}(2003)}]{NOMAD1}
\bibinfo{author}{\bibfnamefont{P.}~\bibnamefont{Astier}} \bibnamefont{et~al.}
  (\bibinfo{collaboration}{NOMAD Collaboration}), \bibinfo{journal}{Phys. Lett.
  B} \textbf{\bibinfo{volume}{570}}, \bibinfo{pages}{19}
  (\bibinfo{year}{2003}).

\bibitem[{\citenamefont{Stockdale et~al.}(1985)\citenamefont{Stockdale, Bodek,
  Borcherding, Giokaris, Lang et~al.}}]{CCFR84}
\bibinfo{author}{\bibfnamefont{I.}~\bibnamefont{Stockdale}},
  \bibinfo{author}{\bibfnamefont{A.}~\bibnamefont{Bodek}},
  \bibinfo{author}{\bibfnamefont{F.}~\bibnamefont{Borcherding}},
  \bibinfo{author}{\bibfnamefont{N.}~\bibnamefont{Giokaris}},
  \bibinfo{author}{\bibfnamefont{K.}~\bibnamefont{Lang}}, \bibnamefont{et~al.},
  \bibinfo{journal}{Z. Phys. C} \textbf{\bibinfo{volume}{27}},
  \bibinfo{pages}{53} (\bibinfo{year}{1985}).

\bibitem[{\citenamefont{Dydak et~al.}(1984)\citenamefont{Dydak, Feldman, Guyot,
  Merlo, Meyer et~al.}}]{CDHS}
\bibinfo{author}{\bibfnamefont{F.}~\bibnamefont{Dydak}},
  \bibinfo{author}{\bibfnamefont{G.}~\bibnamefont{Feldman}},
  \bibinfo{author}{\bibfnamefont{C.}~\bibnamefont{Guyot}},
  \bibinfo{author}{\bibfnamefont{J.}~\bibnamefont{Merlo}},
  \bibinfo{author}{\bibfnamefont{H.}~\bibnamefont{Meyer}},
  \bibnamefont{et~al.}, \bibinfo{journal}{Phys. Lett. B}
  \textbf{\bibinfo{volume}{134}}, \bibinfo{pages}{281} (\bibinfo{year}{1984}).

\bibitem[{\citenamefont{Mahn et~al.}(2012)}]{Mahn:2011ea}
\bibinfo{author}{\bibfnamefont{K.~B.~M.} \bibnamefont{Mahn}}
  \bibnamefont{et~al.} (\bibinfo{collaboration}{SciBooNE, MiniBooNE}),
  \bibinfo{journal}{Phys. Rev.} \textbf{\bibinfo{volume}{D85}},
  \bibinfo{pages}{032007} (\bibinfo{year}{2012}), \eprint{1106.5685}.

\bibitem[{\citenamefont{Cheng et~al.}(2012)}]{Cheng:2012yy}
\bibinfo{author}{\bibfnamefont{G.}~\bibnamefont{Cheng}} \bibnamefont{et~al.}
  (\bibinfo{collaboration}{SciBooNE, MiniBooNE}), \bibinfo{journal}{Phys. Rev.}
  \textbf{\bibinfo{volume}{D86}}, \bibinfo{pages}{052009}
  (\bibinfo{year}{2012}), \eprint{1208.0322}.

\bibitem[{\citenamefont{{D.G. Michael {\it et al.} (MINOS
  Collaboration)}}(2006)}]{MINOSCC1}
\bibinfo{author}{\bibnamefont{{D.G. Michael {\it et al.} (MINOS
  Collaboration)}}}, \bibinfo{journal}{Phys. Rev. Lett.}
  \textbf{\bibinfo{volume}{97}}, \bibinfo{pages}{191801}
  (\bibinfo{year}{2006}).

\bibitem[{\citenamefont{{P. Adamson {\it et al.} (MINOS
  Collaboration)}}(2008)}]{MINOSCC2}
\bibinfo{author}{\bibnamefont{{P. Adamson {\it et al.} (MINOS
  Collaboration)}}}, \bibinfo{journal}{Phys. Rev. D}
  \textbf{\bibinfo{volume}{77}}, \bibinfo{pages}{072002}
  (\bibinfo{year}{2008}).

\bibitem[{\citenamefont{Conrad et~al.}(2013)\citenamefont{Conrad, Ignarra,
  Karagiorgi, Shaevitz, and Spitz}}]{Conrad:2012qt}
\bibinfo{author}{\bibfnamefont{J.~M.} \bibnamefont{Conrad}},
  \bibinfo{author}{\bibfnamefont{C.~M.} \bibnamefont{Ignarra}},
  \bibinfo{author}{\bibfnamefont{G.}~\bibnamefont{Karagiorgi}},
  \bibinfo{author}{\bibfnamefont{M.~H.} \bibnamefont{Shaevitz}},
  \bibnamefont{and} \bibinfo{author}{\bibfnamefont{J.}~\bibnamefont{Spitz}},
  \bibinfo{journal}{Adv. High Energy Phys.} \textbf{\bibinfo{volume}{2013}},
  \bibinfo{pages}{163897} (\bibinfo{year}{2013}), \eprint{1207.4765}.

\bibitem[{\citenamefont{Abazajian et~al.}(2012)}]{sterileneutrinowhitepaper}
\bibinfo{author}{\bibfnamefont{K.~N.} \bibnamefont{Abazajian}}
  \bibnamefont{et~al.} (\bibinfo{year}{2012}), \eprint{1204.5379}.

\bibitem[{\citenamefont{Giunti and Kim}(2007)}]{Giunti:2007ry}
\bibinfo{author}{\bibfnamefont{C.}~\bibnamefont{Giunti}} \bibnamefont{and}
  \bibinfo{author}{\bibfnamefont{C.~W.} \bibnamefont{Kim}},
  \emph{\bibinfo{title}{{Fundamentals of Neutrino Physics and Astrophysics}}}
  (\bibinfo{year}{2007}).

\bibitem[{\citenamefont{Parke and Ross-Lonergan}(2015)}]{Parke:2015goa}
\bibinfo{author}{\bibfnamefont{S.}~\bibnamefont{Parke}} \bibnamefont{and}
  \bibinfo{author}{\bibfnamefont{M.}~\bibnamefont{Ross-Lonergan}}
  (\bibinfo{year}{2015}), \eprint{1508.05095}.

\bibitem[{\citenamefont{Foreman-Mackey
  et~al.}(2013)\citenamefont{Foreman-Mackey, Hogg, Lang, and
  Goodman}}]{foreman-mackey_emcee:_2013}
\bibinfo{author}{\bibfnamefont{D.}~\bibnamefont{Foreman-Mackey}},
  \bibinfo{author}{\bibfnamefont{D.~W.} \bibnamefont{Hogg}},
  \bibinfo{author}{\bibfnamefont{D.}~\bibnamefont{Lang}}, \bibnamefont{and}
  \bibinfo{author}{\bibfnamefont{J.}~\bibnamefont{Goodman}},
  \bibinfo{journal}{Publications of the Astronomical Society of the Pacific}
  \textbf{\bibinfo{volume}{125}}, \bibinfo{pages}{306} (\bibinfo{year}{2013}),
  ISSN \bibinfo{issn}{00046280, 15383873}, \bibinfo{note}{arXiv: 1202.3665},
  \urlprefix\url{http://arxiv.org/abs/1202.3665}.

\bibitem[{\citenamefont{Baker and Cousins}(1984)}]{baker_clarification_1984}
\bibinfo{author}{\bibfnamefont{S.}~\bibnamefont{Baker}} \bibnamefont{and}
  \bibinfo{author}{\bibfnamefont{R.~D.} \bibnamefont{Cousins}},
  \bibinfo{journal}{Nuclear Instruments and Methods in Physics Research}
  \textbf{\bibinfo{volume}{221}}, \bibinfo{pages}{437} (\bibinfo{year}{1984}),
  ISSN \bibinfo{issn}{0167-5087},
  \urlprefix\url{http://www.sciencedirect.com/science/article/pii/0167508784900164}.

\bibitem[{\citenamefont{Feldman and Cousins}(1998)}]{feldman_unified_1998}
\bibinfo{author}{\bibfnamefont{G.~J.} \bibnamefont{Feldman}} \bibnamefont{and}
  \bibinfo{author}{\bibfnamefont{R.~D.} \bibnamefont{Cousins}},
  \bibinfo{journal}{Physical Review D} \textbf{\bibinfo{volume}{57}},
  \bibinfo{pages}{3873} (\bibinfo{year}{1998}), ISSN \bibinfo{issn}{0556-2821,
  1089-4918}, \bibinfo{note}{arXiv: physics/9711021},
  \urlprefix\url{http://arxiv.org/abs/physics/9711021}.

\bibitem[{\citenamefont{Chen and Shao}(1999)}]{chen_monte_1999}
\bibinfo{author}{\bibfnamefont{M.-H.} \bibnamefont{Chen}} \bibnamefont{and}
  \bibinfo{author}{\bibfnamefont{Q.-M.} \bibnamefont{Shao}},
  \bibinfo{journal}{Journal of Computational and Graphical Statistics}
  \textbf{\bibinfo{volume}{8}}, \bibinfo{pages}{69} (\bibinfo{year}{1999}),
  ISSN \bibinfo{issn}{1061-8600, 1537-2715},
  \urlprefix\url{http://www.tandfonline.com/doi/abs/10.1080/10618600.1999.10474802}.

\bibitem[{\citenamefont{Aguilar-Arevalo}(2015)}]{aaaprivate}
\bibinfo{author}{\bibfnamefont{A.}~\bibnamefont{Aguilar-Arevalo}}
  (\bibinfo{year}{2015}).

\bibitem[{\citenamefont{Aguilar-Arevalo et~al.}(2007)}]{miniboonelowe1}
\bibinfo{author}{\bibfnamefont{A.~A.} \bibnamefont{Aguilar-Arevalo}}
  \bibnamefont{et~al.} (\bibinfo{collaboration}{MiniBooNE}),
  \bibinfo{journal}{Phys. Rev. Lett.} \textbf{\bibinfo{volume}{98}},
  \bibinfo{pages}{231801} (\bibinfo{year}{2007}), \eprint{0704.1500}.

\bibitem[{\citenamefont{Aguilar-Arevalo et~al.}(2010)}]{nubarminiosc1}
\bibinfo{author}{\bibfnamefont{A.~A.} \bibnamefont{Aguilar-Arevalo}}
  \bibnamefont{et~al.} (\bibinfo{collaboration}{MiniBooNE}),
  \bibinfo{journal}{Phys. Rev. Lett.} \textbf{\bibinfo{volume}{105}},
  \bibinfo{pages}{181801} (\bibinfo{year}{2010}), \eprint{1007.1150}.

\bibitem[{\citenamefont{Antonello et~al.}(2015)}]{SBNproposal}
\bibinfo{author}{\bibfnamefont{M.}~\bibnamefont{Antonello}}
  \bibnamefont{et~al.} (\bibinfo{collaboration}{LAr1-ND, ICARUS-WA104,
  MicroBooNE}) (\bibinfo{year}{2015}), \eprint{1503.01520}.

\bibitem[{\citenamefont{Goodman and Weare}(2010)}]{goodman_ensemble_2010}
\bibinfo{author}{\bibfnamefont{J.}~\bibnamefont{Goodman}} \bibnamefont{and}
  \bibinfo{author}{\bibfnamefont{J.}~\bibnamefont{Weare}},
  \bibinfo{journal}{Communications in Applied Mathematics and Computational
  Science} \textbf{\bibinfo{volume}{5}}, \bibinfo{pages}{65}
  (\bibinfo{year}{2010}), ISSN \bibinfo{issn}{2157-5452, 1559-3940},
  \urlprefix\url{http://msp.org/camcos/2010/5-1/p04.xhtml}.

\bibitem[{\citenamefont{Earl and Deem}(2005)}]{earl_parallel_2005}
\bibinfo{author}{\bibfnamefont{D.~J.} \bibnamefont{Earl}} \bibnamefont{and}
  \bibinfo{author}{\bibfnamefont{M.~W.} \bibnamefont{Deem}},
  \bibinfo{journal}{Physical Chemistry Chemical Physics}
  \textbf{\bibinfo{volume}{7}}, \bibinfo{pages}{3910} (\bibinfo{year}{2005}),
  ISSN \bibinfo{issn}{1463-9084},
  \urlprefix\url{http://pubs.rsc.org/en/content/articlelanding/2005/cp/b509983h}.

\end{thebibliography}





\end{document}